\documentclass[a4paper]{article}

\usepackage{INTERSPEECH2020}
\usepackage{multirow}
\usepackage{makecell}
\usepackage[flushleft]{threeparttable}
\usepackage{hyperref}
\usepackage{amsmath}
\usepackage{soul,color}
\usepackage{subcaption}
\usepackage{enumitem}
\usepackage{xcolor}

\title{TalkNet: Fully-Convolutional Non-Autoregressive Speech Synthesis Model}
\name{
    Stanislav Beliaev$^1$\textsuperscript{*}\thanks{\textsuperscript{*}Project accomplished during an internship at NVIDIA},
    Yurii Rebryk$^1$,
    Boris Ginsburg\thanks{\textsuperscript{**}Preprint. Submitted to INTERSPEECH.}
}
\address{
 NVIDIA, Santa Clara, USA\\
 $^1$Higher School of Economics, Saint Petersburg, Russia
}
\email{\{stanislavv,bginsburg\}@nvidia.com, y.a.rebryk@gmail.com} 

\begin{document}

\maketitle

\begin{abstract}

We propose TalkNet, a convolutional non-autoregressive neural model for speech synthesis. The model consists of two feed-forward convolutional networks. The first network predicts grapheme durations. An input text is expanded by repeating each symbol according to the predicted duration. The second network generates a mel-spectrogram from the expanded text.
To train a grapheme duration predictor, we add the grapheme duration to the training dataset using a pre-trained Connectionist Temporal Classification (CTC)-based speech recognition model. The explicit duration prediction eliminates word skipping and repeating. Experiments on the LJSpeech dataset show that the speech quality nearly matches auto-regressive models. The model is very compact -- it has 10.8M parameters, almost 3x less than the present state-of-the-art text-to-speech models. 
The non-autoregressive architecture allows for fast training and inference.

\end{abstract}


\section{Introduction}

Neural Network (NN) based models for text-to-speech (TTS) have outperformed both concatenative and statistical parametric speech synthesis in terms of speech quality. They also significantly simplify the speech synthesis pipeline. The traditional TTS pipeline combines multiple blocks: a front end for extracting linguistic features from text, a duration model, an acoustic feature prediction model, and a signal-processing-based vocoder \cite{Taylor2009}. Neural TTS systems typically have two stages. In the first stage, a model generates  mel-spectrograms from text. In the second stage, a NN-based vocoder synthesizes speech from the mel-spectrograms. Most NN-based TTS models have an encoder-attention-decoder architecture \cite{bahdanau2014neural}, which has been observed to have some common problems:
\begin{enumerate}
    \item A tendency to repeat or skip words \cite{Fastspeech2019}, due to attention failures when some subsequences are repeated or ignored. To handle this issue, attention-based models use additional mechanisms to encourage monotonic attention \cite{Tacotron2,DeepVoice3,Taigman2017}.
    \item Slow inference relative to parametric models.
    \item No easy way to control prosody nor voice speed, since the length of the generated sequence is automatically determined by the decoder \cite{Fastspeech2019}.
\end{enumerate}

\begin{figure}[!ht]
  \centering
  \includegraphics[width=0.95\linewidth]{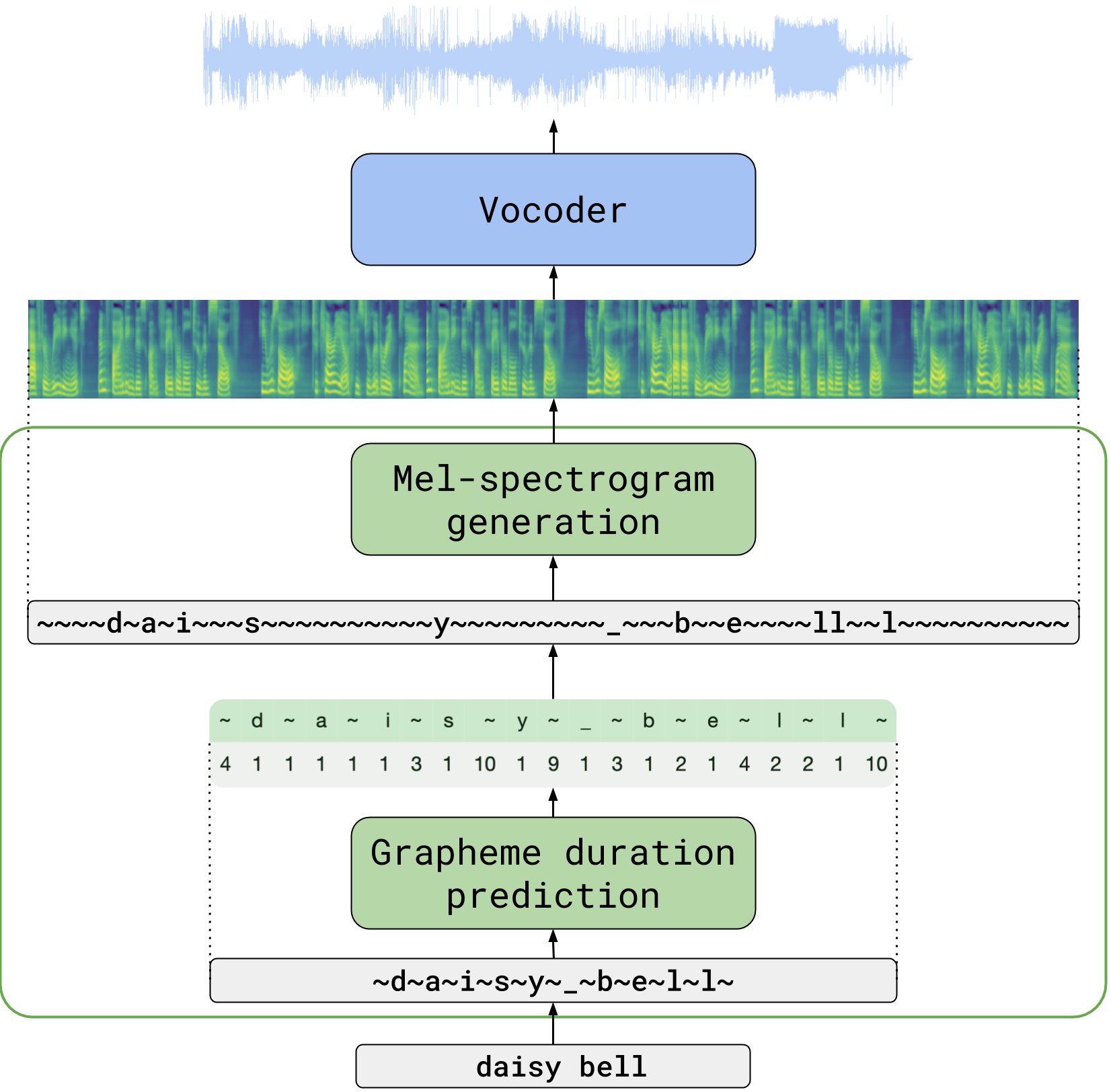}
  \caption{TalkNet converts text to speech, using a grapheme duration predictor, a mel-spectrogram generator, and a vocoder. We use $\sim$ to denote the CTC blank symbol.}
  \label{fig:architecture}
\end{figure}

We propose a new neural TTS model to address these three challenges. The model consists of two convolutional networks. The first network predicts grapheme durations. We expand an input text by repeating each symbol according to the predicted duration. The second network generates mel-spectrograms from an expanded text. Finally, we use the WaveGlow~\cite{Waveglow2019} vocoder to synthesize audio from mel-spectrograms (see Figure~\ref{fig:architecture}).

To train the grapheme duration predictor, we need the ground truth alignment between input characters and audio. A similar alignment problem exists in automatic speech recognition (ASR) which is addressed by using Connectionist Temporal Classification (CTC). CTC marginalizes over the set of all valid alignments. However, if we take the most likely output at each moment, we can use it for alignment between the input audio and the text output. This alignment is not perfect, and it can have errors. We found that if the ASR model is accurate and has a low character error rate (CER), then we can extract a good-enough alignment between text and audio features. We can use this CTC-based alignment to train the model which will predict grapheme durations for the input text.
The explicit grapheme duration predictor replaces attention-based alignment to prevent word skipping and repeating. Experiments on the LJSpeech dataset show that the speech quality for TalkNet is similar to auto-regressive models.

The convolutional structure of both blocks enables parallel training and inference. This structure enables significantly faster inference, has significantly fewer parameters, and can be trained faster than models with similar quality of generated speech, such as FastSpeech~\cite{Fastspeech2019} and Tacotron 2~\cite{Tacotron2}.

\section{Related work}

A typical statistical parametric TTS pipeline has the following stages: grapheme-to-phoneme conversion, a phoneme duration predictor, an acoustic frame-level feature generator, and a vocoder \cite{Taylor2009}. 
Zen et al \cite{Zen2015, Zen2016} proposed a hybrid NN--parametric TTS model, where deep neural networks are used to predict the phoneme duration and to generate frame-level acoustic features. The phoneme duration predictor was trained on Hidden Markov Model (HMM)-based phonetic alignments.  

DeepVoice models \cite{DeepVoice1, DeepVoice2} also adopt the traditional TTS structure, but they replace all components with NNs. To train the phoneme duration predictor, an auxiliary CTC-based model for phonetic segmentation was used to annotate  data with phoneme boundaries. 
Tacotron \cite{Tacotron1, Tacotron2} is an end-to-end NN which takes characters as input and directly outputs the mel-spectrogram. Tacotron 2 uses an encoder-attention-decoder architecture. The encoder is composed from three convolutional layers plus a single bidirectional LSTM. The decoder is a recurrent neural network (RNN) with location-sensitive monotonic attention.

The sequential nature of RNN-based models limits the training and inference efficiency. There has been a number of TTS models without RNNs.
DeepVoice 3 \cite{DeepVoice3} replaces an RNN with a fully-convolutional encoder-decoder with monotonic attention. Switching from RNN to a convolutional neural network (CNN) makes training faster, but the model inference is still auto-regressive. Another end-to-end TTS model, which does not use RNNs,  is  ParaNet \cite{Paranet2019}. ParaNet is a convolutional encoder-decoder with attention. It distills attention from a teacher auto-regressive TTS model. Lastly, Transformer-TTS \cite{TransformerTTS} replaces an RNN-based encoder-decoder with a Transformer-like  attention-only architecture \cite{vaswani2017attention}. 
% Transformer-TTS first converts text to phonemes using a rule-based converter. Using phoneme sequences as input, Transformer-TTS generates mel-spectrograms.  

As with other attention-based models, Tacotron, Transformer-TTS and ParaNet occasionally miss or repeat words \cite{Paranet2019}.
To prevent word skipping and repeating, FastSpeech \cite{Fastspeech2019} proposes a novel feed-forward Transformer-based model, discarding the conventional encoder-attention-decoder structure. FastSpeech uses an explicit length regulator, which expands the hidden sequence of phonemes according to a predicted duration in order to match the length of a mel-spectrogram sequence. The target phoneme duration is extracted from the attention alignment in an external pre-trained TTS model, Tacotron 2. 

\section{System architecture}

TalkNet splits the text-to-spectrogram generation into two separate modules. The first module, the duration predictor, aligns input graphemes in time with respect to the audio features. The second module, the mel-spectrogram generator, produces mel-spectrograms from time-aligned input characters. We use feed-forward CNNs for both modules, so both training and inference are non-autoregressive. This allows for much faster training and inference compared to auto-regressive models. To train the grapheme duration predictor we extracted ground truth alignment from the CTC output of a pre-trained ASR model.  

\subsection{Ground truth grapheme duration construction}

\begin{figure}[!ht]
  \centering
  \includegraphics[width=1.0\linewidth]{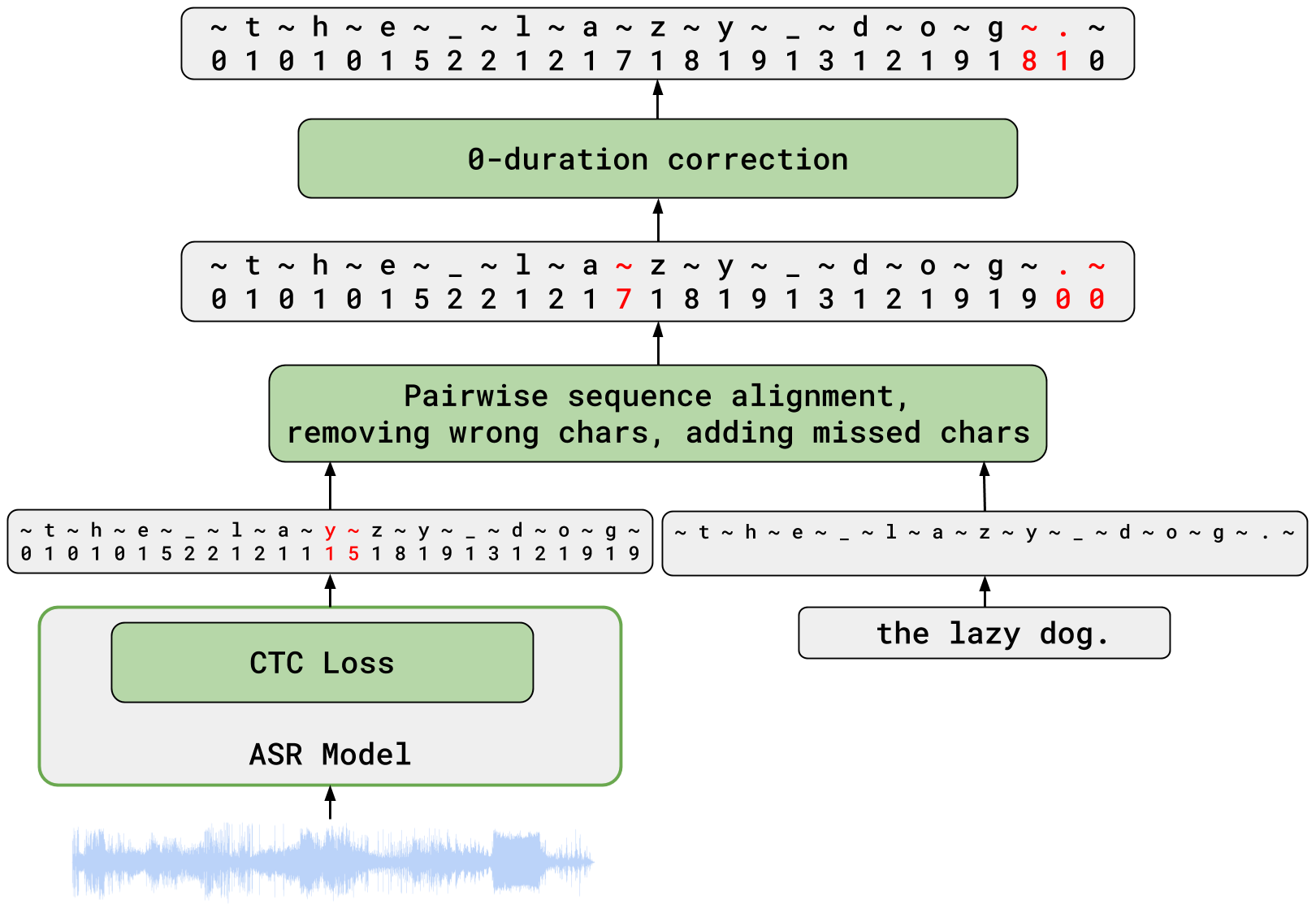}
  \caption{Grapheme duration extraction from CTC output. We use $\sim$ to denote the blank symbol.}
  \label{fig:duration_extraction}
\end{figure}

The central idea behind TalkNet is to use a CTC-based ASR model to extract grapheme alignments. CTC assigns a probability to each of characters from the alphabet, with an auxiliary blank symbol $\sim$. The blank symbol acts as an intermediate state between two neighbouring graphemes, and its duration corresponds to the length of the transition from one character to another. For each time step, we choose the most likely character from the CTC output. 
Because the CTC output is imperfect, we align it with the ground truth text using the \textit{pairwise2} function from the Biopython~\cite{biopython} package. We then remove all the incorrect characters in the CTC output, adding their duration to the nearest blank, and add the missing characters and set their duration to $0$. Next, for all the characters with predicted duration 0,  we set the duration to 1 by subtracting 1 from the near biggest blank to keep the sum of all grapheme durations equal to the length of a mel-spectrogram (see Figure~\ref{fig:duration_extraction}).

To obtain the ground truth grapheme duration we use QuartzNet15x5~\cite{quartznet} with a minor modification: we set the stride in the first convolutional layer to $1$ to make the length of the CTC output equal to the length of the mel-spectrogram. Apart from this, we keep all punctuation intact while tokenizing the text to let CTC handle punctuation alignment as well.
We train QuartzNet on LibriTTS~\cite{libritts}. We achieve a CER of $4.51\%$ on LibriTTS test-clean and $3.54\%$ on the LJSpeech test set. The alignment obtained from CTC is used to train the grapheme duration predictor.

\subsection{Grapheme duration predictor}

\begin{figure}[!ht]
  \centering
  \includegraphics[width=1.0\linewidth]{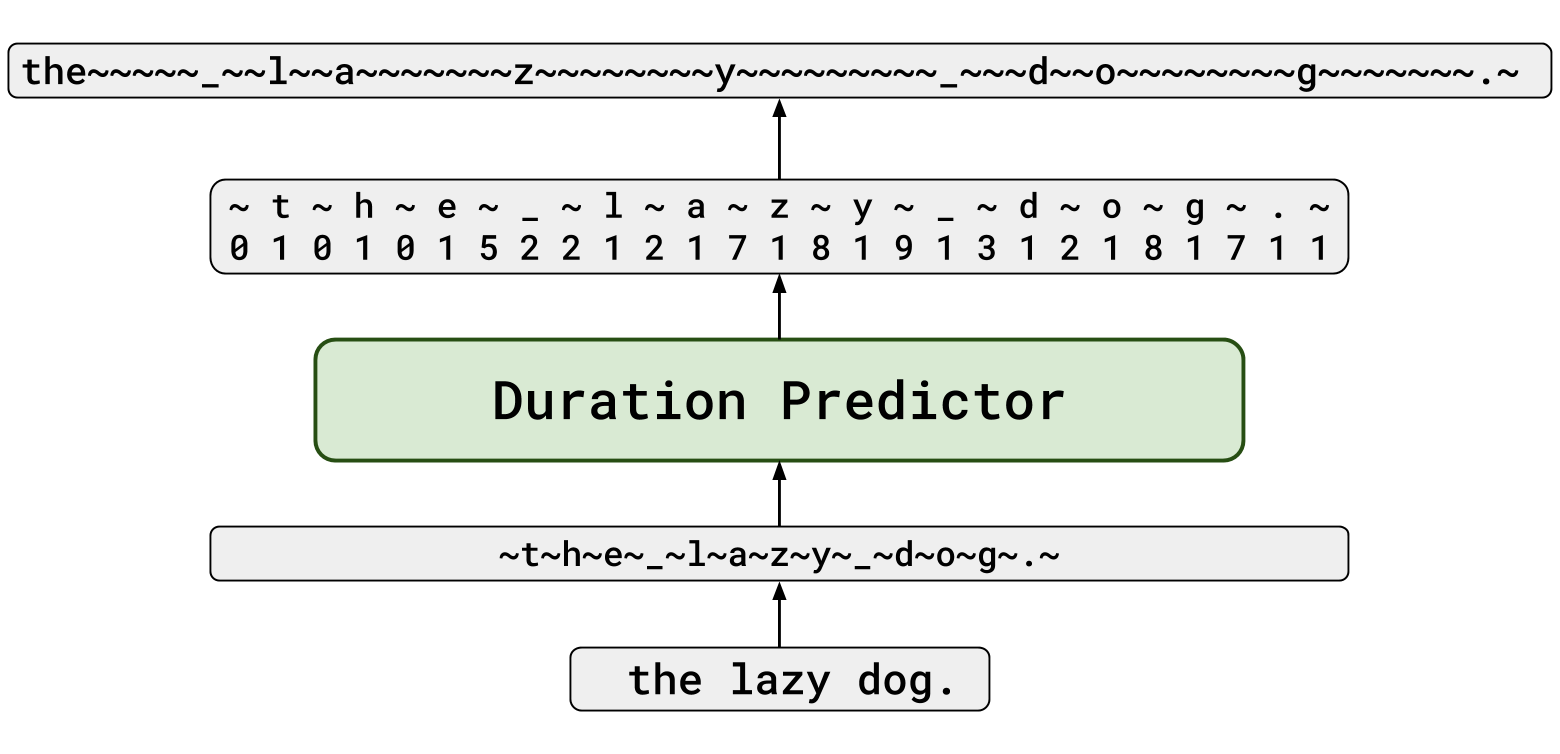}
  \caption{Grapheme duration prediction.}
  \label{fig:durs-logic}
\end{figure}

This model predicts the length of the mel-spectrogram part corresponding to each grapheme in the input including punctuation. First, the grapheme duration predictor inserts a blank symbol $\sim$ between every two input characters. Next, it predicts the duration for each input character. We expand the sequence of input characters by repeating each character according to the predicted duration  (Figure~\ref{fig:durs-logic}).

\begin{figure}[!ht]
  \centering
  \includegraphics[width=0.6\linewidth]{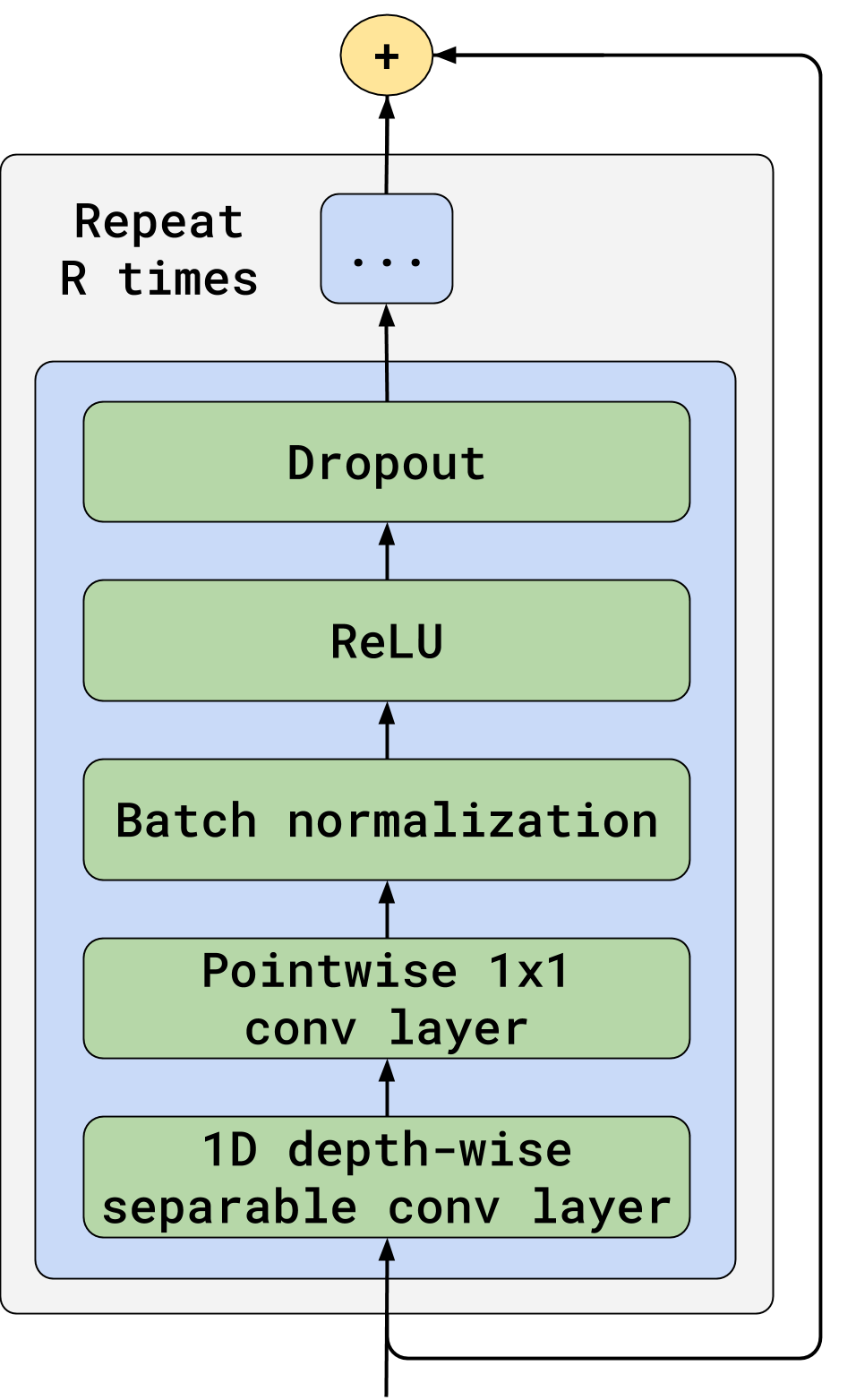}
  \caption{Basic QuartzNet block. Both the grapheme duration predictor and the mel-spectrogram generator are 1D time-channel convolutional networks based on QuartzNet \cite{quartznet}.}
  \label{fig:quartznet_basic_block}
\end{figure}

The grapheme duration predictor model is a 1D time channel separable convolutional NN based on the QuartzNet architecture \cite{quartznet}. 
The model has five residual blocks with five sub-blocks per block. A sub-block consists of a 1D time-channel separable convolution, a $1\times 1$ pointwise convolution, batch normalization, ReLU, and dropout (see Figure~\ref{fig:quartznet_basic_block}).
There are two additional layers: the grapheme embedding layer, and a $1\times1$ convolutional layer before the loss function
(see Table~\ref{tab:durs-model}).

\begin{figure}[!ht]
\centering
\includegraphics[width=.48\linewidth]{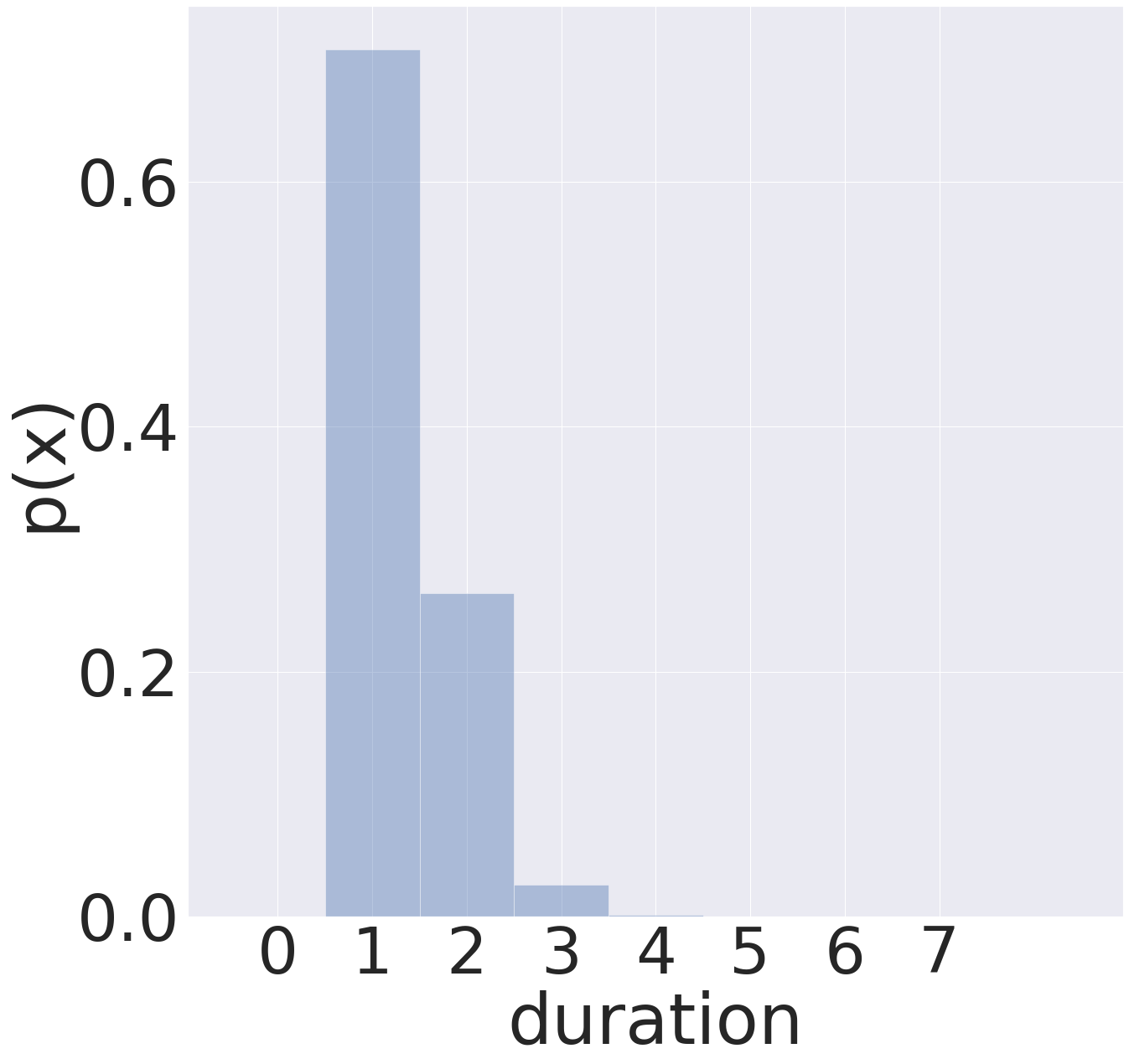}
\includegraphics[width=.48\linewidth]{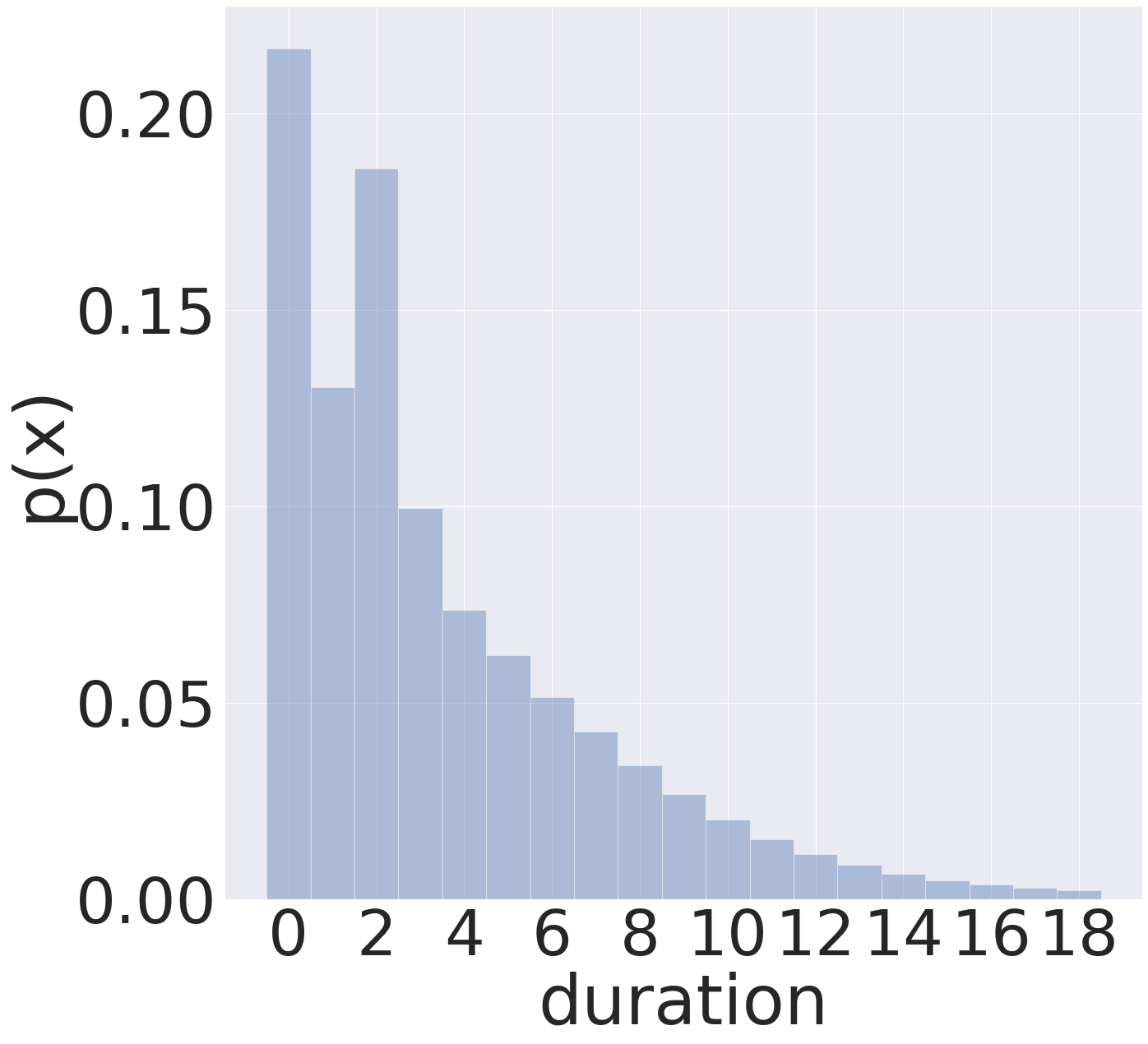}
\caption{The duration distribution for characters (left) and for blanks (right) based on CTC output for the LJSpeech dataset. The maximum duration is $7$ for characters and $493$ for blanks.}
\label{fig:durs-dist}
\end{figure}

We train a duration predictor using $L_2$ loss with logarithmic targets, similar to \cite{Fastspeech2019}. We also tried cross-entropy (XE) loss with each class corresponding to the character duration. We used a log scale for large durations since the grapheme duration distribution has a long tail (Figure~\ref{fig:durs-dist}). Cross entropy has slightly higher accuracy (see Table~\ref{tab:durs-results}). We choose $L_2$ since speech generated with $L_2$ loss received a slightly higher mean-opinion-score (MOS) in our evaluation studies.

\begin{table}[ht!]
\caption{Grapheme duration predictor based on QuartzNet 5x5.}
\label{tab:durs-model}
\centering
\scalebox{0.85}{
\begin{tabular}{c c c c c} 
\toprule
\textbf{Block} &
\textbf{\thead{\# Sub\\Blocks}} &
\textbf{\thead{\# Output\\Channels}} &
\textbf{Kernel Size} &
\textbf{Dropout} \\
\midrule
Embed & 1 & 64  & 1 & 0.0  \\
Conv1 & 3 & 256 & 3 & 0.1  \\
$B_1$ & 5 & 256 & 5 & 0.1  \\
$B_2$ & 5 & 256 & 7 & 0.1  \\
$B_3$ & 5 & 256 & 9 & 0.1  \\
$B_4$ & 5 & 256 & 11 & 0.1 \\
$B_5$ & 5 & 256 & 13 & 0.1 \\
Conv2 & 1 & 512 & 1 & 0.1  \\
Conv3 & 1 & $32$ & 1 & 0.0 \\
\midrule
\textbf{Parameters (millions)} & & & & \textbf{2.3} \\
\bottomrule
\end{tabular}
}
\end{table}

{\renewcommand{\arraystretch}{1.0}
\begin{table}[!ht]
\caption{Duration predictor results on the LJSpeech test set. $P$:~predicted duration, $T$: target duration.
}
\label{tab:durs-results}
\centering
\scalebox{0.85}{
\begin{tabular}{c c c c c c c} 
\toprule
\textbf{Method} &
\textbf{MSE} &
\textbf{Accuracy ($\%$)} &
$\mathbf{|P - T| \leq 1}$ &
$\mathbf{|P - T| \leq 3}$\\
\midrule
$L_2$ & 7.81 & 67.69 & 91.90 & 97.17 \\
XE & 10.46 & 69.42 & 92.90 & 97.40 \\
\bottomrule
\end{tabular}
}
\end{table}
}

\subsection{Mel-spectrogram generator}

The second module generates mel-spectrograms from the expanded text. The mel-spectrogram generator is a 1D convolutional network based on the same QuartzNet architecture. It has nine blocks with five sub-blocks (see Table~\ref{tab:mel-predictor}). The mel-spectrogram generator was trained with a mean square error (MSE) loss.

Instead of allocating a separate embedding for the blank symbol, we use a linear combination of embeddings for the neighboring graphemes. If the blank symbol $\sim$ is located between the  characters $a$ and $b$ and the blank duration is $d$, then the embedding $E$ for the blank symbol located at the distance $t$ from $a$ would be $E(\sim, t) = \dfrac{d+1-t}{d+1} \cdot E(a) + \dfrac{t}{d+1} \cdot E(b)$. 
% Giving $d$ as total blank duration, we choose $c$ to linearly depend on the relative offset $t$ with respect to the left grapheme $a$: $c(t) = \frac{t}{d}$.

\begin{table}[ht!]
\caption{Mel-spectrogram generator based on QuartzNet~9x5.}
\label{tab:mel-predictor}
\centering
\scalebox{0.85}{
\begin{tabular}{c c c c c} 
 \toprule
  \textbf{Block} &
  \textbf{\thead{\# Sub\\Blocks}} &
  \textbf{\thead{\# Output\\Channels}} &
  \textbf{Kernel Size} &
  \textbf{Dropout} \\
 \midrule
Embed & 1 & 256 & 1 & 0.0 \\
Conv1 & 3 & 256 & 3 & 0.0 \\
$B_1$ & 5 & 256 & 5 & 0.0 \\
$B_2$ & 5 & 256 & 7 & 0.0 \\
$B_3$ & 5 & 256 & 9 & 0.0 \\
$B_4$ & 5 & 256 & 13 & 0.0 \\
$B_5$ & 5 & 256 & 15 & 0.0 \\
$B_6$ & 5 & 256 & 17 & 0.0 \\
$B_7$ & 5 & 512 & 21 & 0.0 \\
$B_8$ & 5 & 512 & 23 & 0.0 \\
$B_9$ & 5 & 512 & 25 & 0.0 \\
Conv2 & 1 & 1024 & 1 & 0.0 \\
Conv3 & 1 & 80 & 1 & 0.0 \\
 \midrule
 \textbf{Parameters (millions)} & & & & \textbf{8.5} \\
 \bottomrule
\end{tabular}
}
\end{table}

\section{Training}

\subsection{Dataset}

We use the LJSpeech~\cite{ljspeech} dataset for experiments.
We randomly split the dataset into three sets: $12,500$ samples for training, $300$ samples for validation, and $300$ samples for testing. We lowercased the input text while leaving all punctuation.
We convert ground truth audio to mel-spectrograms through a Short-Time Fourier Transform (STFT) using a 50 ms window size, a 12.5 ms frame hop, and a Hann window.

\subsection{Grapheme duration predictor training}

The NN for grapheme duration was trained using the Adam optimizer with $\beta_1=0.9,\beta=0.999,\epsilon=10^{-8}$, a weight decay of ${10}^{-6}$ and gradient norm clipping of $1.0$. We used a cosine decay learning rate policy starting from $10^{-3}$ and decreasing to $10^{-5}$ with a $2\%$ warmup. We used a batch size of $256$ for one 16GB GPU and scaled learning rate for multi-GPU setups. We trained the grapheme duration predictor for $200$ epochs. Training on one V100 GPU takes approximately $1.3$ hours on one GPU, and approximately $11$ minutes on a DGX1 server with 8 GPUs in mixed precision~\cite{micikevicius2017mixed}.

\subsection{Mel-spectrogram generator training}

As can be seen in Table~\ref{tab:durs-results}, the duration predictor accuracy is around $70\%$, but it covers approximately $92\%$ of classes within an absolute distance of $1$. Mitigating this train/inference mismatch, we used \textit{duration augmentation} to improve model robustness with respect to errors in the duration predictor. For example, we randomly adjust the duration between adjacent characters while keeping the total length unchanged. The change in each character duration is unbiased and proportional to the duration value.

We train the mel-spectrogram generator for $200$ epochs with the same training parameters as above. We used a batch size of $64$ for one 16GB GPU, and scale the learning rate for a multi-GPU setup. Training takes approximately $8$ hours for one V100 GPU and less than $2$ hours for 8 GPUs in mixed precision.

\section{Results}

\subsection{Audio quality}

We conduct the MOS (mean opinion score) evaluation for generated speech using Amazon Mechanical Turk. We compared four types of samples: 1) ground truth speech, 2) ground truth mel-spectrogram converted to speech with WaveGlow, 3) Tacotron 2 + WaveGlow, and 4) TalkNet + WaveGlow. We used NVIDIA's implementation for Tacotron 2 and WaveGlow. We tested $100$ audio samples with $10$ people per sample. The scores ranged from $1.0$ to $5.0$ with a step of $0.5$. TalkNet speech quality comes quite close to Tacotron 2 (see Table~\ref{tab:mos}). 

\begin{table}[!ht]
\centering
\caption{MOS scores with $95\%$ confidence interval}
\label{tab:mos}
\scalebox{0.85}{
    \begin{tabular}{l c} 
    \toprule
    \textbf{Model} &
    \textbf{MOS} \\
    \midrule
    Ground truth speech & $4.31 \pm 0.05$ \\
    Ground truth mel + WaveGlow & $4.04 \pm 0.05$ \\
    Tacotron 2 + WaveGlow & $3.85 \pm 0.06$ \\
    % FastSpeech~\cite{Fastspeech2019} & $\star \pm \star$ \\
    \midrule
    TalkNet + WaveGlow & $3.74 \pm 0.07$ \\
    \bottomrule
    \end{tabular}
}
\end{table}

TalkNet is very robust with respect to missing or repeated words compared to auto-regressive TTS models such as Tacotron 2 or Transformer TTS. We evaluated the  robustness of TalkNet on 50 hard sentences from the FastSpeech paper~\cite{Fastspeech2019} and found that TalkNet practically eliminates missed or repeated words.

\subsection{Inference latency}

In the inference mode, we first insert blank symbols into the tokenized input text between every two characters. The obtained sequence is passed through the grapheme duration predictor. The output of the grapheme duration predictor is then corrected for characters with $0$ duration. The corrected character sequence is expanded with each character repeated according to the predicted duration. The second network generates the mel-spectrogram from the expanded grapheme sequence.
% Finally, we use WaveGlow vocoder to generate audio from mel-spectrogram.

We compare TalkNet inference latency with Tacotron 2 and FastSpeech. We used an internal NVIDIA FastSpeech implementation since the original FastSpeech was not available at the time of evaluation. To measure the latency, we generate mel-spectrograms with a batch size equal to 1 for 2048  samples from the LJSpeech dataset. The average mel-spectrogram length is $520$ frames. We benchmark the latency on one V100 GPU. TalkNet inference is significantly lower than Tacotron 2 and FastSpeech (see Table~\ref{tab:tts-models-lats}). Since TalkNet  does not use an attention mechanism, the inference latency does not depend on the input length.

{\renewcommand{\arraystretch}{1.0}
\begin{table}[!ht]
\caption{TalkNet inference latency for mel-spectrogram generation (without vocoder). The latency was measured with batch size $1$ using a V100 GPU and averaged over 2048 samples from LJSpeech. Latency and Real-Time-Factor (RTF) with $95\%$ confidence interval.}
\label{tab:tts-models-lats}
\centering
\scalebox{0.85}{
\begin{tabular}{l  l r} 
\toprule
\textbf{Model} & 
% \textbf{\thead{\# Batch\\size}} &
\textbf{\thead{Inference\\Latency, s}} &
\textbf{RTF} \\
\midrule
% Transformer TTS~\cite{TransformerTTS} & 1 & $6.735 \pm 3.969$ & $1.48 \pm 0.87$ \\
Tacotron 2~\cite{Tacotron2}  &$0.817 \pm 1\cdot 10^{-2} $ & $7.56 \pm 0.01$ \\
FastSpeech~\cite{Fastspeech2019}  & $0.029 \pm 2 \cdot {10}^{-4}$  & $221.01 \pm 1.75$ \\
\midrule
TalkNet  &  $0.019 \pm 1 \cdot {10}^{-5}$ & $328.65 \pm \ \ 4.76$ \\
%  & 4 &   $0.023 \pm 5 \cdot {10}^{-5}$ & $1048.80 \pm 21.75$ \\
%  & 8 &  $0.037 \pm 4 \cdot {10}^{-4}$ & $1340.09 \pm \ \ 8.90$ \\
\bottomrule
\end{tabular}
}
\end{table}
}

\section{Conclusions}

In this paper, we present TalkNet, a fully convolutional neural speech synthesis  system. The model is composed of two convolutional networks: a grapheme duration predictor and a mel-spectrogram generator. The model does not require another text-to-speech model as a teacher. The ground truth grapheme alignment is extracted from the CTC output of a pretrained speech recognition model.

The explicit duration predictor practically eliminates skipped or repeated words. TalkNet achieves a comparable level of speech quality to Tacotron 2 and FastSpeech. 
The model is very compact. It has only $10.8$M parameters, almost 3x less than similar neural TTS models: Tacotron-2 has 28.2M, and FastSpeech has 30.1M parameters. Training TalkNet takes only around $2$ hours on a server with 8 V100 GPUs. The parallel mel-spectrogram generation makes the inference significantly faster.

% I did a quick correction for grammar. I think Chris also did a large grammar review for the first few sections as well.
% I would avoid the points on fast training, since auto-regressive models are also trained in parallel.

The model, the training recipes, and audio samples will be released as part of the NeMo toolkit \cite{nemo2019}.
% The colab notebook with complete TalkNet inference pipeline can be accessed here: \TODO{LINK}. Audio samples are available at \TODO{LINK.}

\section{Acknowledgments}
The authors thank Jon Cohen, Vitaly Lavrukhin, Jason Li, Christopher Parisien, and Joao Felipe Santos for the helpful feedback and review. 
% The authors thank the NVIDIA AI Applications team for thefeedback and review. .

\bibliography{bibliography}

\end{document}